\documentclass[aps,prb,twocolumn,showpacs,amsmath,amssymb,floatfix]{revtex4}
\usepackage{epsfig}
\usepackage{bm}

\def\be{\begin{equation}}
\def\ee{\end{equation}}
\def\bea{\begin{eqnarray}}
\def\eea{\end{eqnarray}}

\begin{document}

\title{Gap theory of rectification in ballistic three-terminal conductors}

\author{ Andrew N. Jordan$^1$ and Markus B\"uttiker$^2$}
\affiliation{$^1$\ Department of Physics and Astronomy, University of Rochester, Rochester, New York 14627, USA \\
$^2$\ D\'{e}partment de Physique Th\'{e}orique, Universit\'{e} de Gen\`{e}ve,
CH-1211 Gen\`{e}ve 4, Switzerland
}
\date{October 2, 2007}

\begin{abstract}
We introduce a model for rectification in three-terminal ballistic conductors, where the central connecting node is modeled as a chaotic cavity.  For bias voltages comparable to the Fermi energy, a strong nonlinearity is created by the opening of a gap in the transport window.  Both noninteracting cavity electrons at arbitrary temperature as well as the hot electron regime are considered. Charging effects are treated within the transmission formalism using a self-consistent analysis. The conductance of the third lead in a voltage probe configuration is varied to also model inelastic effects.  We find that the basic transport features are insensitive to all of these changes, indicating that the nonlinearity is robust and well suited to applications such as current rectification in ballistic systems.  Our findings are in broad agreement with several recent experiments.
\end{abstract}
\maketitle
There has been recent interest in applying ballistic three-terminal junctions as voltage rectifiers or diodes
in emerging nanoelectric technology.  From an applications point of view, the observed nonlinear $I\!-\!V$ 
curves \cite{SongA,SongB,Linke00,Shor01,Wor01,Haan04,Xu04,Mateos04,Gonzalez04,Wallin04,Wallin06,Feldman07} in these junctions are quite attractive in that the effect persists to room temperature and originates without any special engineering. 
Initial experimental investigation into the issue sought to add the source of nonlinearity in various ways:  The experiment Ref.~\onlinecite{SongA,SongB} placed a triangular obstacle in a cavity in an effort to ``force'' the electrons in one direction or another. For some time it was thought that such enhancements were necessary \cite{Geisel,Haan04} but other discussions \cite{Xu01} pointed to a more general origin of the rectification effects. \cite{Shor01} These discussions\cite{SongA,SongB,Xu01,Geisel} are all based on an application of the transmission approach for multi-terminal conductors \cite{Buttiker86} in the non-linear regime. However, beyond the linear regime application of the transmission approach requires a 
self-consistent treatment.\cite{Buttiker03,Christen96} Indeed, it has been shown that ballistic junctions are very sensitive to side-gating. \cite{Hartmann06}  

Even in the linear regime ballistic four-probe junctions have found applications in Hall micromagentometry \cite{Geim92,Geim98} and scanning Hall probe microscopy.\cite{Oral96,Bending99}  Recent advances include a vector Hall sensor.\cite{Vector03}  Multi-terminal ballistic junctions are also found to be sensitive potentiometers.\cite{Peeters99}

In mesoscopic physics, the properties of ballistic four-probe junctions were originally investigated in the linear transport regime starting with work by Roukes et al. \cite{Roukes87} who found at low temperatures an absence (quenching) of the Hall effect. Different geometries were investigated leading to an enhanced or suppressed Hall effect depending only on the geometry of the Hall cross.\cite{Ford89,Chang89}
Already at He temperatures these effects can be well described by classical trajectories.\cite{Bee88}  Interference effects play a role at much lower temperatures.\cite{Baranger88}
In the non-linear regime, interference effects in chaotic cavities have recently found  interest in connection with the generation of rectification effects. These works examine (predominantly) the second order in voltage term of the I-V-characteristic \cite{Christen96} and demonstrate that interactions lead to deviations from the Onsager
symmetry.\cite{Sanchez04,Spivak04,Polianski06,Zumbuhl06,Leturcq06,Angers07} In comparison, the somewhat extreme conditions of large bias and high temperature envisioned for applications of ballistic structures as rectifiers and diodes are outside the scope of the mesoscopic physics literature and thus require a separate treatment. 

\begin{figure}[b]
\epsfxsize=7cm
\epsfbox{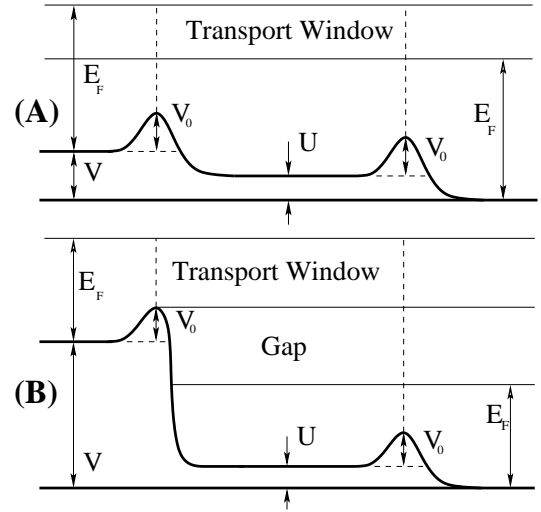}
\caption{Energy landscape for the low $\bf (A)$ and high $\bf (B)$ bias case. In case $\bf (A)$ transport is approximately linear, and the current-carrying electrons have energies in the transport window.  In case $\bf (B)$ the electrical bias $V$ is sufficiently large that an energy gap opens between the transport window and the filled Fermi sea of the right lead.  The energy gap is responsible for the strong nonlinearity in the transport characteristics.}
\label{high}
\end{figure} 

The purpose of this article is to present a simple model of classical rectification in ballistic chaotic cavities.  We here take a minimalist approach to the problem, and make only the following two assumptions:  ({\it i}) Transport between lead and cavity via the quantum point contacts (QPC) is ballistic so the Landauer formula applies, and ({\it ii}) The mean level spacing in the cavity is much smaller than the charging energy of the cavity, and therefore charge neutrality of the cavity under non-equilibrium conditions is imposed.   From these two assumptions we develop our model, and demonstrate that when the applied voltage is comparable to the Fermi energy, a strong nonlinearity develops.  Furthermore, we demonstrate that this behavior is insensitive to the details of the model, showing that this mechanism is generic and robust.

\section{Minimal Model}
We now describe the model in detail.  A ballistic cavity is connected via three QPCs to bulk leads.  We assume that the energy barriers of the QPCs (see Fig. 1{\bf A}) are specified by a potential energy $V_0$.  Electrical bias $V$ is applied across two of the leads.  When the third lead is in the voltage probe configuration, there are two basic DC-transport characteristics:  the dependence of the potential $V_p$ on the third lead (voltage probe), as a function of $V$, and the $I\!-\!V$ curve through the left and right leads.  When the third lead is electrically fixed to be the mid-point voltage between the left and right lead, there are then the three currents through the leads, linked by current conservation.   

In the voltage probe situation, an even simpler model that captures the basic physics is to truncate the third lead by pinching off the third contact, and to consider the dependence of the internal cavity potential $U$ versus $V$.  We will first work out this simplest case analytically in detail, and present only the numerical results for more realistic extensions of this model.

For a chaotic cavity the occupation function is isotropic \cite{Langen97,Blanter00} and the main theoretical task is to find its dependence on energy. The use of a chaotic cavity gives the results of our minimal model a degree of universality which is absent in ballistic junctions with short geometry-dependent trajectories.

\begin{figure}[tbh]
\epsfxsize=7cm
\epsfbox{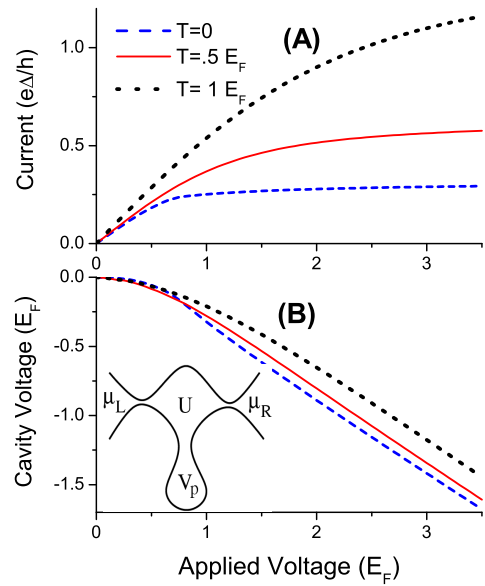}
\caption{{\bf (A)} Current $I$ between left and right lead is plotted versus applied bias $V$ in units of $E_F$ for different temperatures. We take $V_0=.2 E_F$, so the cross-over is around $E_F-V_0=.8 E_F$.  Increasing temperature raises the internal potential $U$ for large applied bias. {\bf (B)} Cavity potential $U-V/2$ plotted versus applied bias $V$ in units of $E_F$,  for different temperatures.  Inset of {\bf (B)}: Considered geometry.  External bias is applied across left and right lead, $\mu_L = \mu_R + e V$.  The cavity voltage is measured with the help of a third voltage probe.}
\label{fig2}
\end{figure} 

{\it Low bias limit, $V < E_F - V_0$}.---  We first assume zero temperature and elastic scattering.  The current going into the cavity from both leads is given by the Landauer formula
\be
I_\alpha = \int dE\, j_\alpha(E) =  (e/h) \int dE\, T_\alpha(E) (f_\alpha -f_C),
\label{current}
\ee
where $\alpha = L,R$, $j(E)$ is the energy-resolved current, $T(E)$ is the energy-dependent transmission, and $f_{L,R,C}$ are the occupation functions of each region. We assume that the lead occupation functions are completely specified by Fermi functions with a single potential: $E_F + V$ and $E_F$.
The large energy scales involved imply that the transmission of the QPCs may be treated semi-classically, so only the coarse energy-dependence is kept:
\be
T_\alpha(E) = \frac{(E - E_\alpha)}{\Delta_\alpha} \Theta(E - E_\alpha),
\label{trans}
\ee 
where we use a linear interpolation (valid for energies larger than the conductance quantization scale) and $E_L = V+V_0,\ E_R = U+V_0$ are the minimal energies required for carriers to pass through the QPC's (see also Fig. 1{\bf B}).  The energy scales $\Delta_\alpha$ characterize how open the contacts are.  The limit $\Delta\rightarrow \infty$ corresponds to a closed contact.  Other energy dependence (such as the semi-classical 3/2 transmission law) leads to similar physics. 

Imposing energy-resolved current conservation $j_L(E) + j_R(E)=0$, allows us to solve for the cavity occupation as a function of energy:
\be
f_{\rm low} = \frac{E - V- V_0}{2E - 2 V_0 - V -U},
\label{fc}
\ee
where $E_F < E < E_F + V$.  Between $U$ and $E_F$, the cavity occupation is 1.
We can now calculate the charge in the cavity, 
\be
Q = \sum_{\alpha=L,R,p} C_\alpha (V_\alpha-U) = e D \int dE f_C(E),
\label{Q}
\ee
where $D$ is the (constant) density of states, $V_\alpha$ is the potential of lead $\alpha$, and $C_\alpha$ is the capacitance linking the central cavity to terminal $\alpha$.  For simplicity, we focus on the realistic case of reasonably large cavities, where the cavity mean level spacing, $D^{-1}$, is much smaller than the charging energy of the cavity, $e^2/C$.  In this case, Eq.~(\ref{Q}) is a charge neutrality condition, and the left hand side may be replaced by the equilibrium charge on the cavity, $Q_0 = e D E_F$.  Inserting Eq.~(\ref{fc}) into (\ref{Q}), we find
\bea
\label{charge1}
Q/(e D) &=& E_F = E_F - U + V/2 \\ &-& [(U-V)/4] \log \left(\frac{2 E_F - V- U- 2 V_0}{2 E_F - U +V - 2 V_0}\right),
\nonumber
\eea
giving a self-consistent (transcendental) equation for the unknown potential $U$, that may be solved numerically.

Notice that the logarithm in Eq.~(\ref{charge1}) originates from keeping the energy dependence of the transmission, and integrating over the full energy range.  Indeed, if we simply neglect the logarithmic correction, we recover the usual linear result, $U=V/2$.  In Fig.~2{\bf (B)} we plot this solution for $V < E_F - V_0$ (the $T=0$ curve only).
In order to take out the usual linear behavior and focus on the nonlinearity, we plot $U-V/2$  (all other plots of cavity or probe potential will reflect this convention).

{\it High bias limit, $V > E_F - V_0$.}---In this range of parameters, the large applied bias lifts the energy barrier of the left QPC up above the Fermi level of the right contact (see Fig.~1{\bf (B)}). This opens up a gap in the transport window, corresponding to an energy range too low for the left current carriers to fill.  Here
\be
f_{\rm high} = \begin{cases}  1 & U <E<E_F, \\ 0 & E_F < E< V+V_0, \\ f_{\rm low} &  V_0 + V < E < E_F + V.
\end{cases}
\ee
Inserting this to Eq.~(\ref{Q}), we find the equation
\bea
Q/(e D) &=& E_F = E_F - U + (E_F -V_0)/2 \\ &-& [(U-V)/4] \log \left(\frac{2 E_F - V- U- 2 V_0}{2 E_F - U +V - 2 V_0}\right). \nonumber
\eea

We observe that if the logarithmic contribution in Eq. (7) is neglected, we obtain 
$U=(E_{F} - V_{0})/2$, so $U$ goes to a constant as $V$ continues to increase.  This explains why Fig.~3{\bf (A),(C)} as well as the experimental data\cite{Wallin06}  (which plot $U-V/2$) show transitions from a flat dependence to a shifted line with slope -1/2 at large bias $V$.   The energy scale $E_F - V_0$ is the cross-over point.
The numerical solution of $U$ as a function of $V$ is given in Fig.~2{\bf (B)} for $V > E_F - V_0$ ($T=0$ curve only).  Note also the spatial inversion symmetry, $V\rightarrow -V, x \rightarrow -x$, for a symmetric geometry gives the (trivially) symmetric negative voltage behavior as found in recent experiments.\cite{Wallin06} For GaAs, the Fermi energy is around $E_F \sim 20\ meV$, which is of the same order of magnitude as room temperature.

{\it Finite temperature.}---
At finite temperature, again imposing energy-resolved current conservation and solving for the cavity occupation, we find the well-known result
\be
f_C = (T_L f_L + T_R f_R)/(T_L + T_R),
\ee
but this only applies for $E>V+V_0$.  For lower energies, current can flow only from the right lead (or not at all), so here $f_C = f_R$, the equilibrium Fermi function. 

Calculating the total charge in the cavity as before, we find
\bea
Q / (e D)&=&  \int_{U}^{V_0+V}dE\, f_R + \int_{V_0+V}^\infty dE\, \frac{f_L T_L + f_R T_R}{T_L +T_R}\nonumber  \\ &=& T \log [1+\exp(E_F/T)].
\eea
This gives a self-consistent equation for the cavity potential $U$, as a function of $V,T$.  The results are shown in Fig.\ 2{\bf (B)} for different values of temperature $T$.   Higher temperature tends to increase the cavity potential.

It is also interesting to look at the $I\!-\!V$ curve for transport between the left and right lead.  
From Eq.~(\ref{current}), we can now find the current as a function of applied voltage, now that we know the internal potential $U$:
\bea
I_L &=& \frac{e}{h} \int_{V+V_0}^\infty  dE \,T_L (f_L-f_C)=  \\
&=&\frac{e}{h} \int_{V+V_0}^\infty  dE\, \frac{T_L T_R}{T_L+T_R} (f_L-f_R),
\eea
where the transmission $T_{R}$ of Eq.~(\ref{trans}) depends of the cavity potential $U$.
Fig.~2{\bf (A)} shows the $I-V$ curve as temperature is varied.  Higher temperature tends to increase current and smooth the transition.   The current is an antisymmetric function under voltage reversal.

\begin{figure}[htb]
\epsfxsize=8.5cm
\epsfbox{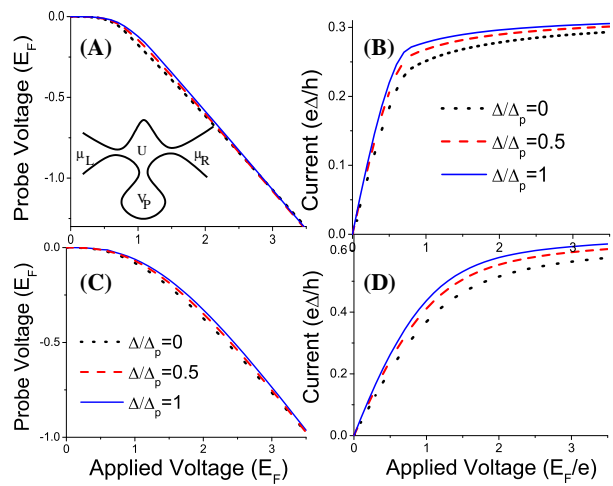}
\caption{ Rectification characteristics as the width of the voltage probe ($\Delta/\Delta_p$) is varied  for $T=0$ {\bf (A), (B)}, and $T=.5 E_F$ {\bf (C), (D)}.  As in Fig. 2, $V_0=.2 E_F$, so the cross-over is around $E_F-V_0=.8 E_F$.
As the temperature increases, the cross-over is more smooth for both the probe voltage {\bf (C)} and the current {\bf (D)}.  When the probe is turned on from closed ($\Delta/\Delta_p=0$) to open ($\Delta/\Delta_p=1$), the $T=0$ curves are essentially indifferent, while the $T=.5 E_F$ curves have slightly more variation.}
\label{multi-temp}
\end{figure}

\section{Voltage probe model}
We now open up the third lead on the structure.  Adding the third lead does two things:  First, it allows a realistic probe of the cavity voltage, which is the essence of the rectification effect, and second it allows a way of treating inelastic processes in a phenomenological way.   This latter effect occurs because high energy carriers can enter the probe, and be reinjected at lower energy.\cite{Buttiker88}  In reality, charge redistribution in energy occurs not only due to the voltage probe but through various inelastic scattering mechanisms.\cite{Heiblum89,Kaya07} In our model the probe can be turned on and off with a coupling parameter. 

The equations are similar to those of the previous section:  Now there are three energy-resolved currents, $j_L, j_R, j_p$ where $j_p$ is the energy resolved current entering the voltage probe.  We require 
\begin{itemize}
\item{({\it i}) Energy-resolved current conservation, $j_L + j_R + j_p=0$.}
\item{({\it ii}) Charge neutrality, $Q = e D \int dE \, f_C(E) = e D E_F$. }
\item{({\it iii}) A new condition:  No net current flow in or out of the voltage probe, $\int dE \, j_p(E) =0$.   }
\end{itemize}
These equations determine both cavity and probe voltage.  We take $j_p(E) = (e/h)[(E-V_0-V_p)/\Delta_p] (f_p - f_C)$. 
The conductance of the probe lead can be smoothly turned on and off by varying $\Delta/\Delta_p$ from 0 (the off configuration), up to $\infty$ where it dominates over the other leads.

{\it Low bias limit, $V < E_F -V_0$.}---
Solving for the cavity occupation from ({\it i}), we find
\be
f_C = \begin{cases}  1 & U <E<E_F, \\ f_C^- & E_F < E< E_F+V_p, \\ f_C^+ &  E_F + V_p < E < E_F + V,
\end{cases}
\ee
where
\bea
f_C^- &=& \frac{E-V_0-V + (\Delta/\Delta_p) (E-V_0-V_p)}{2E -2 V_0 -V -U + (\Delta/\Delta_p) (E-V_0 -V_p)}, \nonumber \\
f_C^+ &=& \frac{E-V_0-V}{2E -2 V_0 -V -U + (\Delta/\Delta_p) (E-V_0 -V_p)}.
\eea
The first equation from ({\it ii}) is given by
\be
Q/eD = E_F -U + \int_{E_F}^{E_F+V_p} dE\, f_C^- +  \int_{E_F+V_p}^{E_F+ V} dE\, f_C^+ = E_F.
\ee
The second equation from ({\it iii}) is then
\begin{align}
&\int_{E_F + V_p}^{E_F +V}  \frac{dE\,(E-V_0-V_p)(E-V_0-V)}{2E -2V_0 -V -U +(\Delta/\Delta_p)(E-V_0-V_p)} \nonumber \\
&= \int_{V_0+V_p}^{E_F + V_p}  \frac{ dE\,(E-V_0-V_p)(E-V_0-U)}{2E -2V_0 -V -U +(\Delta/\Delta_p)(E-V_0-V_p)}.
\end{align}
As before, these integrals may be expressed as logarithms, but the equations are again transcendental, so they need to be solved numerically.

{\it High bias limit, $V > E_F -V_0$.}---
Here also, the main feature is the presence of a gap in the transport window.  There are four energy windows, two of which are nontrivial:  $E \in [U, E_F, E_F +V_p, V+V_0, E_F + V]$.
\begin{itemize}
\item{Filled region: Between $E \in [U, E_F]$ every state is occupied, so $f_C=1$.}
\item{Low energy region:  In the range $E \in [E_F, E_F + V_p]$, the voltage probe can inject carriers into the cavity.  Solving for the occupation in this range from energy resolved current conservation, we find,
\be
f_C^{LE} = \frac{E - V_p -V_0}{E -V_p -V_0 + (\Delta_p/\Delta) (E-U-V_0)}.
\ee}
\item{Gap region: Between $E \in [E_F+V_p, V+V_0]$, the left electrons are injected at too high an energy to fill this region, so $j_L=0$, while the energy of the right and probe electron is too low to fill it:
$j_R = (e/h)[(E-U-V_0)/\Delta] (0 - f_C)$, and $j_p = (e/h)[(E-V_p-V_0)/\Delta_p] (0 - f_C)$.  Therefore $f_C=0$, giving the gap.}
\item{High energy region:  In the range $E \in [V+V_0, V+E_F]$, there is injection from the left, and drain to the right.  Current conservation yields
\be
f_C^{HE} = \frac{E - V -V_0}{2E -2 V_0 -V - U + (\Delta/\Delta_p) (E-V_p-V_0)}.
\ee
Notice $\Delta, \Delta_p$ have switched places from the low energy region.}
\end{itemize}
The first equation from ({\it ii}) is given by
\bea
Q/eD &=& E_F = E_F -U  \\
&+& \int_{E_F}^{V_p+E_F} dE\, f_C^{LE}+ 0  +  \int_{V_0 + V}^{E_F+ V} dE\, f_C^{HE}.  \nonumber
\eea
The second equation from ({\it iii}) is given by
\begin{align}
&\int_{E_F}^{E_F +V_p} \frac{dE\, (E-V_p-V_0)(E-V_0-U)}{E -V_0 -U +(\Delta/\Delta_p)(E-V_0-V_p)}\\
&= \int_{V_0+V_p}^{E_F + V_p} \frac{dE\, (E-V_p-V_0)(E-V_0-V)}{2E -2V_0 -V -U +(\Delta/\Delta_p)(E-V_0-V_p)}.
\end{align}

{\it Finite temperature.}---The analysis is somewhat simpler at finite temperature, simply because there are fewer energy regions to keep track of.   Now there are three energy regions to attend to: $E \in [U, V_0+V_p, V_0+V, \infty]$.
Current conservation gives three possible answers for the cavity occupation $f_C$, depending on what energy region we are considering,
\be
f_C = \begin{cases}  f_R & U <E<V_0+V_p, \\ f_{\rm int} & V_0+V_p < E< V_0 + V, \\ 
f_{\rm high} &  V_0 + V < E < \infty.
\end{cases}
\label{occupation}
\ee
where
\bea
f_{\rm int} &=& \frac{ T_p f_p + T_R f_R}{T_p+T_R}, \\
f_{\rm high} &=&  \frac{ T_p f_p + T_L f_L +T_R f_R}{T_p+T_L+T_R}.
\eea
This gives the charge in the cavity as 
\bea
Q/eD &=& \int_{U}^{V_p+V_0} dE\, f_R 
+ \int_{V_0 + V_p}^{V_0+ V} dE\, f_{\rm int},\label{chargec} \\
&+& \int_{V_0 + V}^{\infty} dE\, f_{\rm high}  
=  T \log [1+\exp(E_F/T)].\nonumber
\eea
The no-net-current probe condition ({\it iii}) then reads
\bea
\int dE\, T_p (f_p-f_C) &=& 0 + \int_{V_0 + V_p}^{V_0+ V} dE\, T_p [f_p - f_{\rm int}] \\
&+&  \int_{V_0 + V}^{\infty} dE\, T_p [f_p - f_{\rm high}] =0.\nonumber
\eea
Taking $f_p$ to be a Fermi function with unknown energy $V_p$, these two equations may be numerically solved to give $U$ and $V_p$ as a function of the parameters $T, V, \Delta, \Delta_p$.

The results are given in Fig.~3.  We see that the basic rectification features remain, and changing the model parameters does not alter the basic picture, indicating a robust voltage rectification effect.

\section{Current rectification}
Current rectification occurs whenever there is a net DC current produced by an external AC voltage source.
We now consider this situation in the three-terminal geometry, where the AC voltage signal has a frequency slower than the RC time of the cavity.  This situation may be analyzed by investigating DC transport with chemical potentials $\mu_L = V/2, \mu_R = -V/2, \mu_p = 0$, and how the probe current depends on $V$.  When $V\rightarrow -V$, the left and right leads will switch roles, but because of the reflection symmetry in the problem, the finite probe current produced by the strong nonlinearity will remain unaltered.  Under repeated sign changes the system will sustain a net DC-current from the probe out into the left and right leads.   To be consistent with the previous results, we add $V/2$ to all potentials, and define all currents as positive when they enter the cavity.

It is instructive to estimate what is the theoretical upper limit of the speed of a ballistic rectifier.  This is controlled by the $RC$-time ($\tau_{RC}$) of the cavity which controlles the relaxation of cavity charge:  if the external AC frequency is slower than $\tau_{RC}$ then current rectification will occur because the cavity has time to establish a (nonequilibrium) steady state, while in the opposite limit, rectification is expected to not occur.   Recent measurements for the capacitance of a mesoscopic cavity found $C \sim 1 f F$. \cite{glattli}  Taking on the order of 10 open channels gives a resistance $R \sim 1 k\Omega$. We then estimate the charge relaxation time to be $\tau_{RC} \sim 10^{-12} s$.   These parameters give rectification on a Terahertz scale.

Making use of previous results, everything is the essentially the same, except that the probe voltage is now fixed, $V_p = V/2$, and the current through the probe needs to be calculated.  The cavity occupation $f_C(E)$ written in Eq.~(\ref{occupation}) is the same, as is the charge in the cavity Eq.~(\ref{chargec}), but now with $V_p = V/2$ in both (\ref{occupation},\ref{chargec}).  These conditions set the potential $U$.  We find that the current through the left lead is
\be
I_L=\frac{e}{h} \int_{V+V_0}^\infty  dE\, \frac{T_L [T_p (f_L-f_R)+ T_R(f_L-f_R)]}{T_L+T_R+T_p},
\label{il}
\ee
the current through the right lead is 
\bea
I_R&=&\frac{e}{h}\int_{V+V_0}^\infty  dE\, \frac{T_R [T_p (f_R-f_p)+ T_L(f_R-f_L)]}{T_L+T_R+T_p}\nonumber \\
&+&\frac{e}{h}\int_{V_0+V/2}^{V+V_0}  dE\, \frac{T_R T_p (f_R-f_p) }{T_R+T_p},
\label{ir}
\eea
and the current through the probe is 
\bea
I_p&=&\frac{e}{h}\int_{V+V_0}^\infty  dE\, \frac{T_p [T_L (f_p-f_L)+ T_L(f_p-f_R)]}{T_L+T_R+T_p}\nonumber \\
&+&\frac{e}{h}\int_{V_0+V/2}^{V+V_0} dE\, \frac{T_R T_p (f_p-f_R)}{T_R+T_p}.
\label{ip}
\eea

We can now use the charge equation to find the potential $U$, put this into Eqs.~(\ref{il},\ref{ir},\ref{ip}), and find the currents in all leads.  The results are shown in Figs.~4 and 5 varying temperature and the width of the probe contact.
At small bias, the $I-V$ curves are linear in both the left and right lead, with no current passing through the probe lead.  This is the expected situation for linear transport.  For larger voltages, as the probe lead is gradually opened, the current through the left lead is essentially unchanged, but the current through the probe and right lead both increase.  This is the direct analogue of the probe voltage $V_p$ and cavity voltage $U$ following the lower right voltage in the probe configuration.

\begin{figure*}[tpb]
\epsfxsize=17cm
\epsfbox{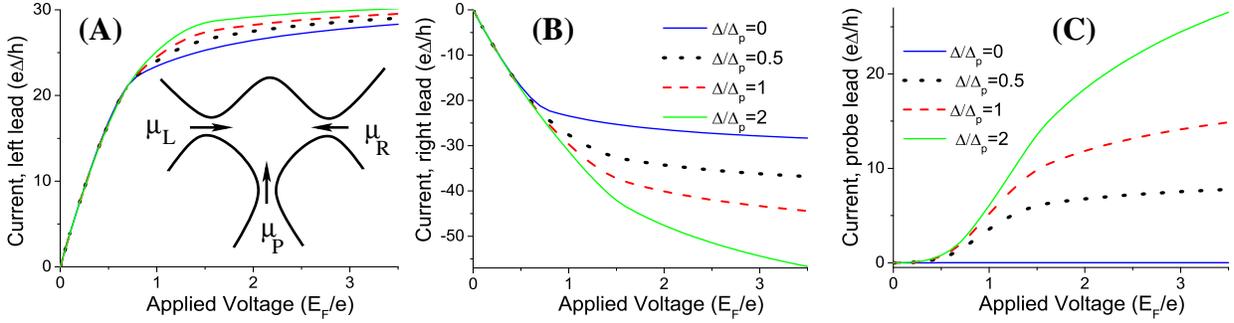}
\caption{ Current rectification effects in each of the three leads for $T=0$. Different curves vary the opening of the probe lead.  For small applied bias, the $I-V$ curves ${\bf (A), (B)}$ start out linear, while the probe lead ${\bf (C)}$ carries no current.  At large bias, the probe lead begins to carry current (so long as it is open, $\Delta/\Delta_p>0$), and this current comes primarily by increasing the current carried by the right lead (the one with low-energy carriers).  Current from the three leads sum to zero.  Clearly, the influence of the probe lead plays a much stronger role here than in Fig.~3.   Inset of {\bf (A)}:  Geometry of the current-rectification set-up. The left, probe and right lead have applied bias $\mu_L = V$,  $\mu_p = V/2$, $\mu_R = 0$.
 }
\label{multi-temp1}
\end{figure*} 
\begin{figure*}[tpb]
\epsfxsize=17cm
\epsfbox{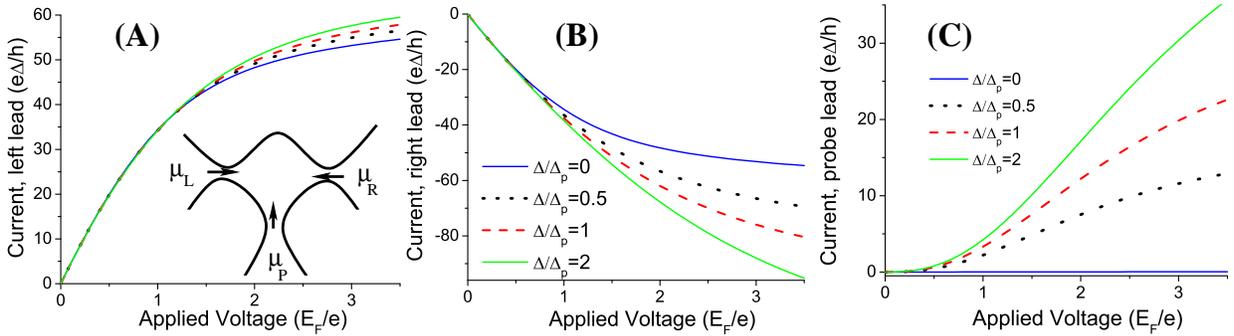}
\caption{Same as Fig.~4, for the case where $T= .5 E_F$.  Apart from smoothing the transitions a little, finite temperature also increases the overall scale of the effect.}
\label{multi-temp2}
\end{figure*} 
\begin{figure*}[tpb]
\epsfxsize=17cm
\epsfbox{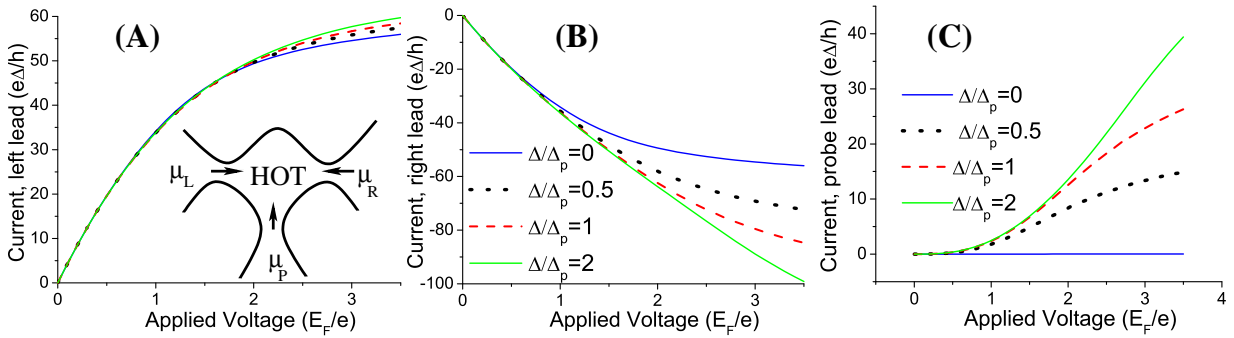}
\caption{Current rectification for the left ${\bf (A)}$, right ${\bf (B)}$, and probe ${\bf (C)}$ lead, using the hot electron model, for $T=.5 E_F$.  Despite the fact that the energy distribution of the cavity electrons is completely altered from the noninteracting case, the curves are still very similar to Fig.~5.
}
\label{hotcombo}
\end{figure*} 

{\it Hot electron regime}.---One weakness of the above analysis is that the gap produced in the energy window is unrealistic at high temperatures.  As electrons collide with one another and phonons, they will redistribute themselves in energy when any inelastic processes are introduced.  One essential point that must be demonstrated is that the nonlinearity we have discovered is not fragile to a reshuffling of electrons in the energy space.  To this end, we will now consider the ``hot-electron regime'', an effective model of electron transport when there is conservation of both charge and energy currents.\cite{Nagaev95,Pilgram03}  In the limit where electron-electron interactions are very strong, $\tau_{ee} < \tau_C$, the cavity comes to a local nonequilibrium steady state, described by a Fermi function with two parameters, $\mu_c$ and $T_c$.  These parameters are determined self-consistently by imposing current conservation and energy conservation for transport through the cavity.  The two constraints may be written as
\be
\int dE \sum_\alpha j_\alpha(E) =0, \qquad  \int dE \sum_\alpha  E j_\alpha(E) =0.
\ee
Once the parameters $\mu_c$ and $T_c$ are found as a function of the tunable parameters of the system, the current through all leads may be found (in the current rectifier mode), or the voltage of the probe may be found (in the voltage probe mode).

The current rectification results are shown in Fig.~6 for the hot electron regime.  We note that the trends in the data are the same, and the only discernible difference is a slight change of shape in some of the curves.  This indicates that the model described here is robust, and not sensitive to changes in the cavity occupation function.

\section{Conclusions}
We have proposed a model of ballistic rectification for a three-terminal geometry.  The model is minimal in the sense that we have only included the most important effects, and therefore should be considered a benchmark theory, rather than designed to predict detailed experimental features.  The most important feature of our model is a cross-over from a weak to a strong nonlinearity when the bias voltage is comparable to the Fermi energy.  This effect has already been observed in experiments.\cite{Wallin06}  The origin of the strong nonlinearity is the opening of a gap in the transport window.  This happens when the applied bias elevates the left QPC energy barrier above the Fermi energy of the right contact.  We have demonstrated that varying parameters in the model, as well as considering inelastic effects do not alter the basic features of the model.  Taken together, this theory indicates that three-terminal ballistic cavities provide robust rectification that may be used in the development of ballistic nonlinear elements such as rectifiers and diodes.

\section*{Acknowledgment}
A.N.J. wishes to thank Marc Feldman and Yonathan Shapir for stimulating discussions about ballistic rectification. M.B. acknowledges the support of the Swiss National Science Foundation, the Swiss Center for Excellence MaNEP and the European STREP project SUBTLE.

\end{document}